\documentclass{article}

\usepackage{epsfig} 
\newlength{\abstractwidth} 
\newcommand{\onefigure}[2]{\begin{figure}[htbp] 
\begin{center}\leavevmode\epsfbox{#1.eps}\end{center}\caption{#2\label{#1}} 
\end{figure}} 
\newcommand{\twofigures}[3]{\begin{figure}[htdp]
\centering
\leavevmode\epsfxsize=2in\epsfbox{#1.eps}
\leavevmode\epsfxsize=2in\epsfbox{#2.eps}
\caption{\small {\em #3}\label{#1}}
\end{figure}}
\newcommand{\threefigures}[4]{\begin{figure}[htdp]
\centering
\leavevmode\epsfxsize=2in\epsfbox{#1.eps}
\leavevmode\epsfxsize=2in\epsfbox{#2.eps}
\leavevmode\epsfxsize=2in\epsfbox{#3.eps}
\caption{\small {\em #4}\label{#1}}
\end{figure}}
\renewcommand{\thefootnote}{\fnsymbol{footnote}} 
\renewcommand{\thanks}[1]{\footnote{#1}} 
\newcommand{\starttext}{ 
\setcounter{footnote}{0} 
\renewcommand{\thefootnote}{\arabic{footnote}}} 
\renewcommand{\theequation}{\thesection.\arabic{equation}} 
\newcommand{\be}{\begin{equation}} 
\newcommand{\bea}{\begin{eqnarray}} 
\newcommand{\eea}{\end{eqnarray}} 
\newcommand{\beq}{\begin{equation}} 
\newcommand{\ee}{\end{equation}} 
\newcommand{\eeq}{\end{equation}}

\def\ba{\begin{eqnarray}} 
\def\ea{\end{eqnarray}}

\def\14{{1\over4}} 
\def\12{{1 \over 2}}

\def\ds{\partial_{\sigma}} 
\def\h3{h^{3\over 2}} 
 
\def\ds{de Sitter Space} 
\def\la{\langle} 
\def\ra{\rangle} 

\newcommand{\tab}{\hspace{5mm}}

\begin{document} 
\renewcommand{\theequation}{\thesection.\arabic{equation}} 
\begin{titlepage} 
\begin{figure}[ht] 
\begin{center} 
\epsfig{file=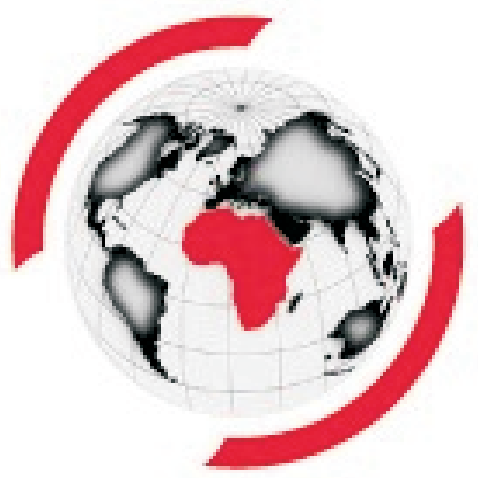,height=4cm} 
\end{center} 
\end{figure} 

\bigskip 
\rightline{SU-ITP 02/07}
\rightline{MIT-CTP 3250} 
\rightline{STIAS-02-003}
\rightline{hep-th/0202163} 

\bigskip\bigskip\bigskip\bigskip 

\centerline{\Large \bf {Is There Really }} \centerline{\Large 
\bf {a de Sitter/CFT Duality }} 
\bigskip\bigskip 
\bigskip\bigskip 
\begin{center} 
{\large Lisa Dyson\footnote{
Permanent address, Center for Theoretical Physics, Massachusetts Institute of Technology, Cambridge, MA   02139},  }
{\large James Lindesay\footnote{
Permanent address, Computational Physics Laboratory, Howard University, Washington, DC  20059},  }
{\large Leonard Susskind\footnote{
Permanent address, Department of Physics, Stanford University, Stanford, CA 94305-4060}
} \\
Stellenbosch Institute for Advanced Study,   Stellenbosch, South Africa 
\end{center}
\bigskip\bigskip 
\begin{abstract} 
In this paper a de Sitter Space version of Black Hole 
Complementarity is formulated which states that an observer in \ds 
\ describes the surrounding space as a sealed finite temperature 
cavity bounded by a horizon which allows no loss of information. 
We then discuss the implications of this for the existence of 
boundary correlators in the hypothesized dS/cft correspondence. We 
find that dS complementarity precludes the existence of the 
appropriate limits. We find that the limits exist only in 
approximations in which the entropy of the \ds \ is infinite. The 
reason that the correlators exist in quantum field theory in the 
\ds \ background is traced to the fact that horizon entropy is 
infinite in QFT. 
\medskip 
\noindent 
\end{abstract} 
\end{titlepage} 
\starttext \baselineskip=17.63pt \setcounter{footnote}{0}

\setcounter{equation}{0} 
\section{ The Complementarity Principle}

\tab Recent evidence suggests that we may live in a space-time that will 
asymptotically 
tend to \ds . If this is so, it is important to understand how 
quantum gravity should be formulated in such a geometry. 
Since \ds \ has an event horizon many of the questions 
that confused theorists about the quantum theory of black holes 
become relevant to cosmology. Perhaps the most important lesson we have 
learned 
from black hole quantum mechanics concerns the complementary 
way that different observers describe events in the black hole environment. 
That together with the Holographic principle, the UV/IR connection and 
the counting of black hole microstates is providing a new paradigm 
for the quantum mechanics of horizons. It is therefore natural to try 
to apply them to \ds .

According to the Principle of Black Hole Complementarity \cite{stretch} the 
horizon of a black hole may be regarded by an external observer as 
an impenetrable thermal membrane which can absorb, thermalize and 
reemit all information. The principle also says that a freely 
falling observer encounters nothing special at the horizon. The 
principle has recieved strong support from the study of gravity in 
AdS space and its equivalence to the boundary conformal field 
theory. 
The horizon of a \ds \ is structurally very similar to that of a 
black hole. A static patch of \ds \ is described by the metric 
\be 
ds^2 = R^2 \left [(1-r^2)dt^2 -(1-r^2)^{-1} dr^2 -r^2 d^2 \Omega 
\right ]. 
\label{static} 
\ee 
In this case the horizon surrounds the observer.

It is well known that the static patch has an entropy given by 
\be 
S={Area \over 4G} ={4 \pi R^2 \over 4G}. 
\label{entropy} 
\ee 
The proper temperature at $r=0$ is given by 
$T={1 \over 2 \pi R}$. More generally the local proper temperature is given 
by 
\be 
T(r)={1 \over 2 \pi R \sqrt{1-r^2}}. 
\label{localtemp} 
\ee 
Note that as in the black hole case the local proper temperature formally 
diverges near the horizon.

The analog of the Black Hole Complementarity Principle for de Sitter space can be expressed as follows: An observer 
in \ds \ sees the surrounding spacetime as a finite closed cavity bounded by a horizon. The cavity is described
by a thermal ensemble at coordinate temperature $1/{2 \pi}$. As for any closed system information can never be 
lost but can be scrambled or thermalized at the hot horizon. 
Also as in the black hole case, a freely falling observer  will experience nothing out of the ordinary when 
passing through the horizon. These complementary descriptions of the horizon comprise the Complementarity Principle for \ds . 

Next let us consider the attempt by Strominger \cite{dscft} to formulate a 
holographic 
description of \ds . The proposal is inspired by the AdS/CFT duality and is 
based on the fact that the 
symmetries of \ds \ act on the asymptotic boundary of \ds \ as conformal 
mappings. In particular time-translation 
in the static patch is a dilatation that preserves a point $p$ on the 
boundary. 
Ordinarily this dilatation invariance 
would allow us to construct a local field at $p$. Take any operator in the 
static patch at time $t$. 
This operator does not have to be a local bulk operator. Now use time 
translation to translate it back in 
time toward $t= - \infty$. From the boundary point of view this is a 
dilation which should shrink the support 
of the operator to a point. 
One of the key assumptions of the dS/CFT correspondence is that the matrix 
elements of such 
an operator behave like 
\be 
\exp{\gamma t} 
\label{gamma} 
\ee 
where the real part of $\gamma$ is positive. Factoring off this exponential 
leaves a set of matrix elements 
that are assumed to define local field operators in the boundary CFT. In 
what follows it will be shown that 
the assumption of eq.~(\ref{gamma}) must be wrong and that no local field is 
defined by the above procedure. 
\setcounter{equation}{0} 
\section{Correlations in QFT} 
\tab As an example we will follow Bousso, Maloney and Strominger who consider a 
massive scalar field $\phi$ in $2+1$ dimensional 
\ds . Straightforward study of the wave equation shows that correlation 
functions behave like 
eq.~(\ref{gamma}) with 
\be 
\gamma=H \pm i \sqrt{m^2-H^2} 
\label{gvalue} 
\ee 
where $H$ is the Hubble parameter proportional to $1/R$. 
As a particular example consider the two-point function $\langle 
\phi(t)\phi(t')\rangle$ 
where both points are evaluated at the spatial point $r=0$. 
Eq.~(\ref{gamma}) together with time translation 
invariance in the static patch imply that the correlation function behaves 
like 
\be 
\langle \phi(t)\phi(t')\rangle \sim \exp{-\gamma |t-t'|}. 
\label{xdecay} 
\ee 
Evidently the correlator exponentially tends to zero with large time 
separation in the static patch. 
\setcounter{equation}{0} 
\section{Correlations for Systems of Finite Entropy} 
\tab The exponential decay of correlations has an interesting 
explanation from the point of view of the thermal cavity picture. 
It's a general property of correlation functions in thermal equilibrium that 
they 
exponentially die off with large time. Typically the coefficient in the 
exponential 
is some dissipation coefficient. The \ds \ correlation functions are simply 
reflecting 
the phenomena of dissipation in a thermal bath. However, as we shall see, 
this is only true if 
the thermal bath has infinite entropy. For example this would be the case 
for an infinite heat bath. 
For a closed system of finite entropy the behavior is much more complicated 
at asymptotic 
times. The essential point is that any quantum system with finite thermal 
entropy must have a discrete spectrum. 
This is because the entropy is essentially the logarithm of the number of 
states per unit energy. 
Of course in many circumstances including classical physics, the entropy arises from an integral 
over a continuum of states. However it is generally understood that the true information theoretic
entropy of such systems contains an additive infinity. 
Thus if we believe that the entropy is finite and that it has the usual 
statistical meaning, 
the spectrum must be discrete. 
Let us now consider a general finite closed system described by a thermal 
density matrix and a 
thermal correlator of the form 
\be 
F(t)=\la A(0) A(t) \ra = {1\over Z}Tr e^{-\beta H} A(0) e^{iHt}A(0) 
e^{-iHt}. 
\label{F} 
\ee 
By finite we simply mean that the spectrum is discrete and the entropy finite. 
Inserting a complete set of (discrete) energy eigenstates gives 
\be 
F(t)={1\over Z} \sum_{ij}e^{-\beta E_i}e^{i(E_j-E_i)t}|A_{ij}|^2. 
\label{foft} 
\ee 
For simplicity we will assume that the operator $A$ has no 
matrix elements connecting states of equal energy. This means 
that the time average of $F$ vanishes.

Let us now consider the long time average of $F(t)F^*(t)$. 
\be 
L=\lim_{{T \to \infty}} {1 \over 2T} \int_{-T}^{+T} dt F(t) F^*(t) 
\ee 
Using eq.~(\ref{foft}) it is easy to show that the long time average is 
\be 
L={1\over Z^2} \sum_{ijkl}e^{-\beta(E_i+E_k)}|A_{ij}|^2 |A_{kl}|^2 
\delta_{(E_j-E_l+E_k -E_i)}. 
\label{lta} 
\ee 
where the delta function is defined to be zero if the argument is non-zero 
and $1$ if it is zero. 
The long time average $L$ is obviously non-zero and positive. Thus it is not 
possible for the correlator $F(t)$ 
to tend to zero as the time tends to infinity and the limits required by the 
dS/CFT correspondence can not exist. 
The value of the long time average for such finite systems can be estimated 
and it is typically of the order of some power of
$e^{-S}$ where $S$ is the entropy of the system. This observation allows 
us to understand why it 
tends to zero in the (bulk) QFT approximation. In studying QFT in the 
vicinity of a horizon it is well known that 
the entropy is UV divergent. This is due to the enormous number of short 
wave length modes near the horizon. 
Any phenomenon which crucially depends on the finiteness of horizon entropy 
will be gotten wrong by the 
approximation of QFT in a fixed background. 
How exactly do the correlations behave in the long time limit? The most 
probable answer is not that they uniformly 
approach constants given by the long time averages. The expected behavior is 
that they fluctuate chaotically. 
A large fluctuation which reduces the entropy by amount $\Delta S$ has 
probability $e^{-\Delta S}$. Thus 
we can expect large fluctuations in the correlators at intervals of order 
$e^{S}$. These fluctuations are closely related 
to the classical phenomenon of Poincare recurrences. In the 
appendix to this paper a numerical study of such correlations 
is presented for finite systems with random matrix Hamiltonians. 
It is found that the large time behavior of correlators is 
chaotic "noise" with the long time average given by 
\be 
L\sim e^{-S}. 
\label{exp} 
\ee

Does this mean that an exact dS/CFT correspondence can not exist? This is 
not entirely clear. Let's suppose that 
the CFT is something like a Euclidean version of a gauge theory and that the 
correspondence is similar to the 
AdS/CFT case. Quantities associated with the bulk would be described by 
nonlocal objects such as Wilson 
loops of finite size; the larger the Wilson loop the further the 
corresponding point from the past boundary. 
A correlator at two times would be related to the product of concentric 
Wilson loops. 
Now in the usual case the local 
gauge invariant operators such as the energy momentum tensor can be obtained 
in terms of shrinking Wilson loops. 
The argument given in this 
paper is that the limit in which one of the loops shrinks to zero can not 
exist but instead must fluctuate in 
endless Poincare recurrences. Can a gauge theory behave this way and can it 
form a representation of the 
conformal group? Perhaps but it would certainly be a very unfamiliar kind of 
theory. 
Another point concerns the implications for string theory in \ds . If 
boundary correlators made any sense they 
would be the natural candidates for computables in string theory much in the 
same way that string theory defines 
observables in AdS. The non-existence of these correlators raises the 
question of what objects string theory can 
define in \ds  \cite{witten}. At the moment there is no answer to this question. 
Finally we should point out that the magnitude of the chaotic 
effects that we have discussed is of order $e^{-S}=e^{-A/4G}$ 
where $A$ is the horizon area of \ds . This has the form of a 
non-perturbative effect in the gravitational coupling. Maldacena 
\cite{juan} 
has discussed very similar effects for eternal black holes in 
AdS and has suggested that they arise from non-perturbative sums 
over space-time topology. This may also be possible in \ds . 
\section{Acknowledgements} 
\tab The authors would like to thank Bernard Lategan and Hendrik Geyer for their 
kind hospitality 
at Stellenbosch University and the Stellenbosch Institute for Advanced Study 
in South Africa where this work was done. 
\setcounter{equation}{0} 
\section{Appendix} 
\tab As an illustration of the late times behavior of systems with 
finite entropy, we will construct a model and numerically explore 
its behavior at large times. We will study the thermal correlators 
of operators A given by the functional form 
of a scalar field $ \phi $ that would behave for large times as in equations 
\ref{gamma} and \ref{gvalue} if the energy spectrum were continuous.  The forms $A_{ij}$ are
then directly related to a Fourier transform of this field, with the energy
differences $ E_{i}-E_{j} $ as its argument, and $\gamma$ as a parameter.

\tab We model 
the finite entropy system by constructing a finite square matrix 
of random elements with a normal (Gaussian) distribution and using 
the eigenvalues of this matrix as the energy levels of the thermal 
system. The standard deviation of the Gaussian distribution 
reflects in the density of the spacing of these energy levels. We 
will assume that the thermal bath has entropy given by $S=-Tr{\rho}log{\rho}$ for the 
thermal density matrix ${\rho}$. The entropy is varied by varying the temperature. 
The time dependence of the absolute 
value of the correlator as given in equation 3.2 can then be 
examined. 
\bigbreak As the entropy increases, the amplitude of 
significant recurrences within a given time frame was seen to 
decrease, as illustrated in the first set of diagrams (Figures \ref{gr1}
a,b,and c) for matrix sizes N=5, 10, and 20 respectively. 
\threefigures{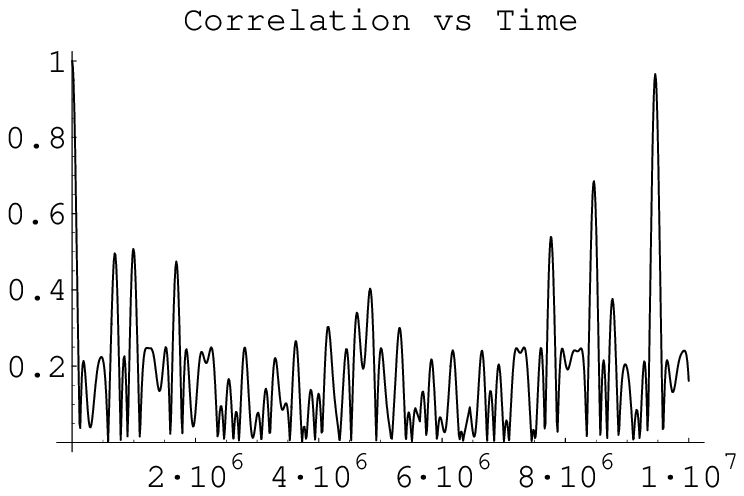}{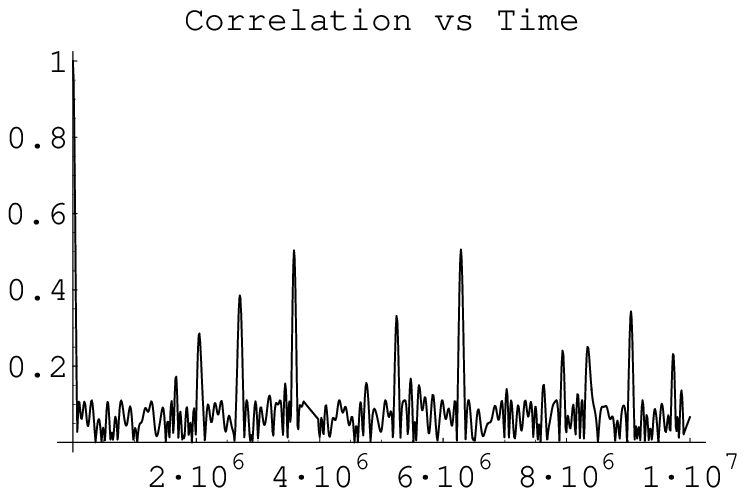}{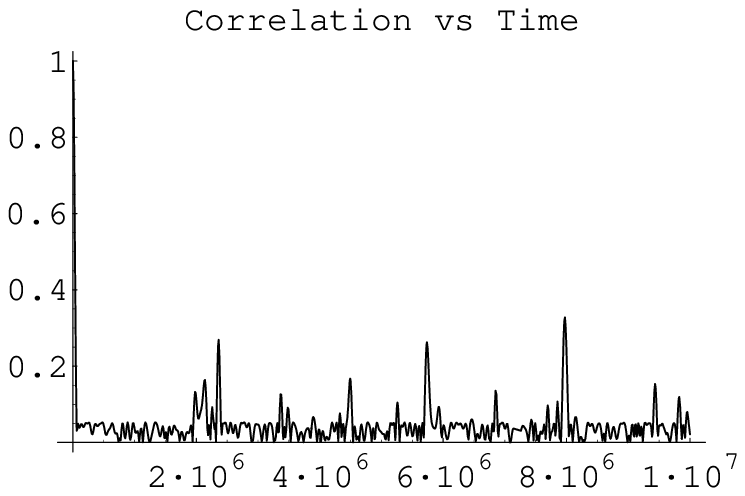}{Large time behavior of correlation functions for (a) N=5, (b) N=10 and (c) N=20}
The time scale for recurrence of significant 
fluctuations above a fixed absolute magnitude was seen to increase roughly 
proportionally to e$^{S}$. This is demonstrated in the next set of 
graphs (Figures \ref{gr4} and \ref{gr6}).   
\twofigures{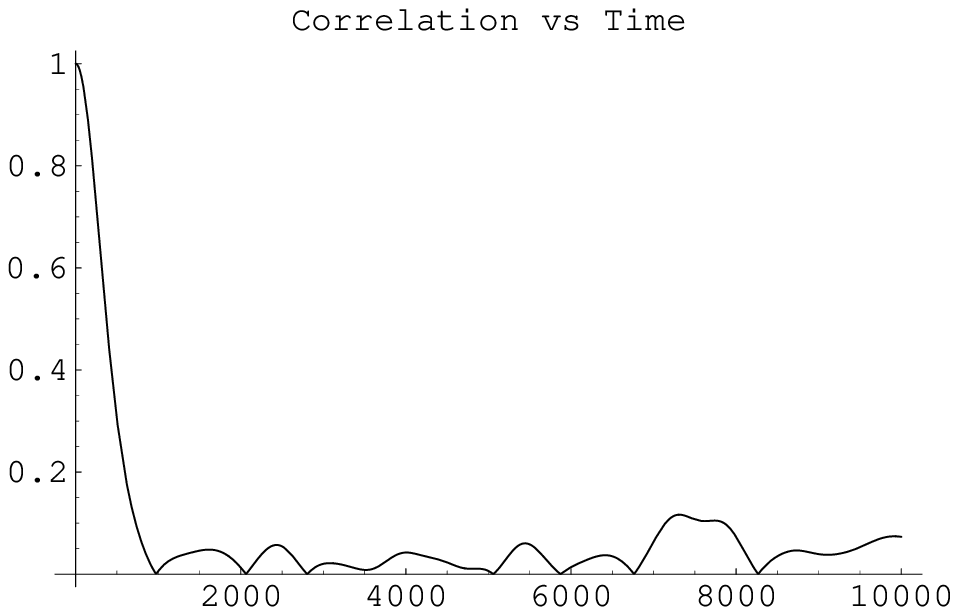}{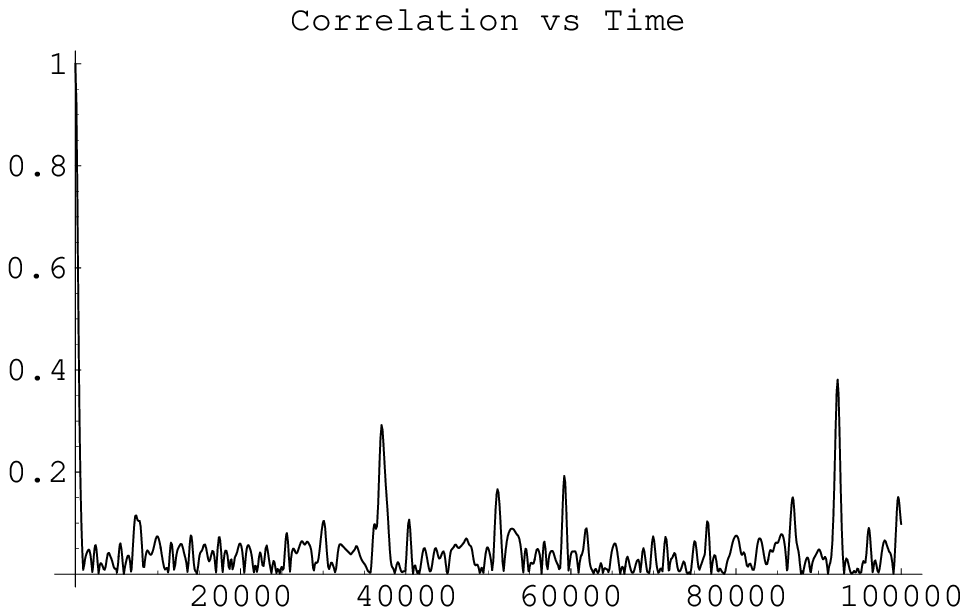}{Correlation recurrences for N=20, showing initial falloff and fluctuations}
\twofigures{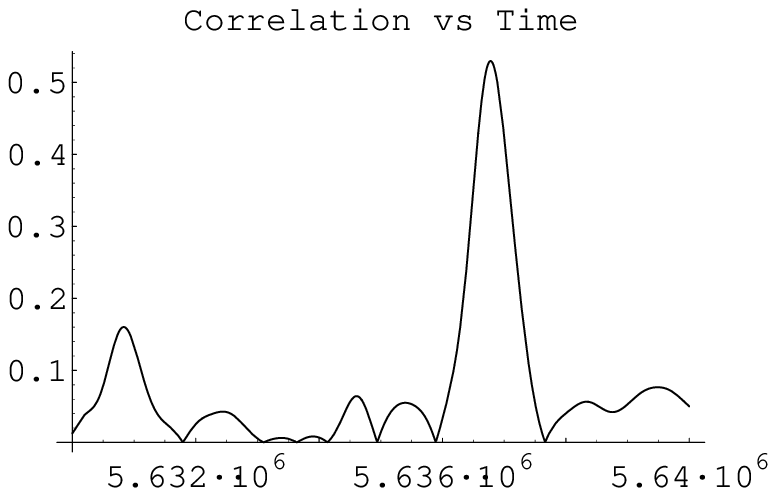}{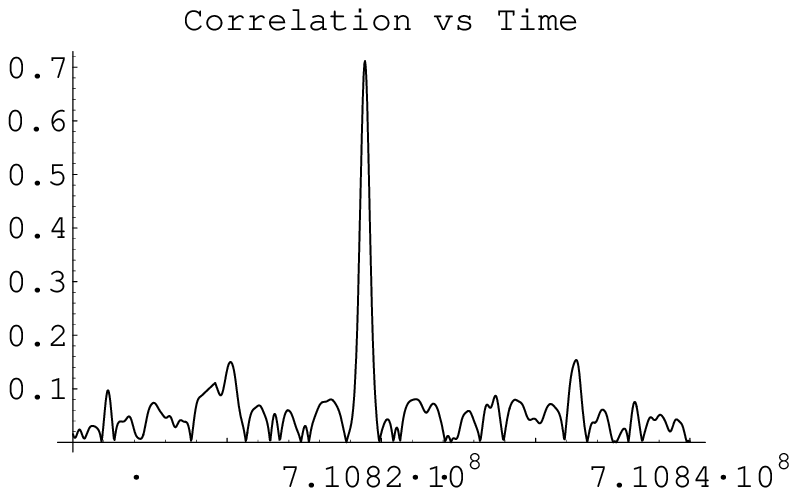}{Correlation recurrences for N=20: (a) 50\%, (b) 70\%}
These graphs for matrix size N=20 demonstrate that fluctuations of 
order 0.4 are seen to occur on time scales of roughly 10$^{4}$ 
units, fluctuations of order 0.5 on time scales of 10$^{6}$, and 
fluctuations of order 0.7 on time scales of 10$^{8}$. Such fixed 
order fluctuations are found to have recurrence times that 
increase very rapidly with the size of the fluctuation. The next 
graph (Figure \ref{gr8}) demonstrates the recurrence time for significant 
fixed size fluctuations as a function of the size of the 
fluctuation for a system with N=10 levels.   
\onefigure{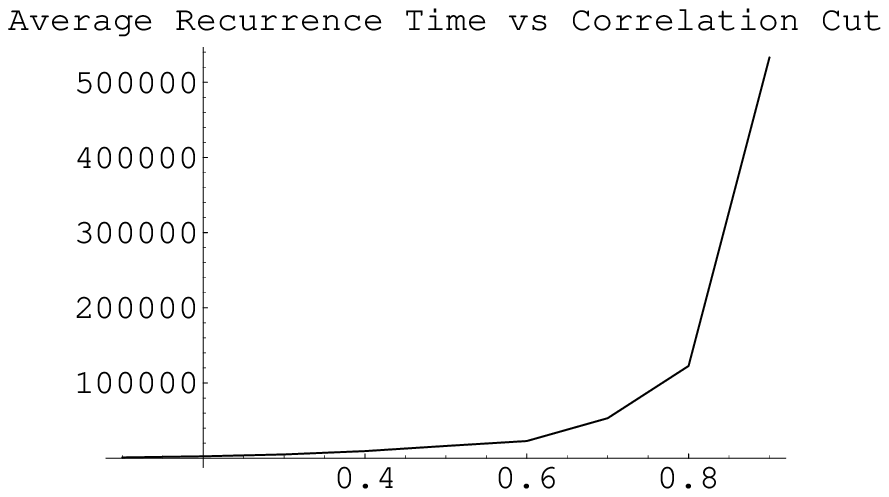}{Recurrence time as a function of absolute size of fluctuations} 

The numerical dependence of these late time fluctuations upon 
the various time scales was explored in an attempt to determine 
the scaling variables. Assuming that all of the external 
parameters (such as de Sitter scale H and scalar mass m) are held 
fixed, there are several time scale parameters in the model which might be 
relevant to the recurrences; for example, the entropy timescale e$^{S}$, the 
inverse range in energy eigenvalues defined for the thermal system 
$1/{\Delta}E$, and the inverse of the average energy spacing 
between the levels N$/{\Delta}E$, along with various combinations 
of these time scales.  The functional behavior found is illustrated in Figure \ref{gr9f}.
\onefigure{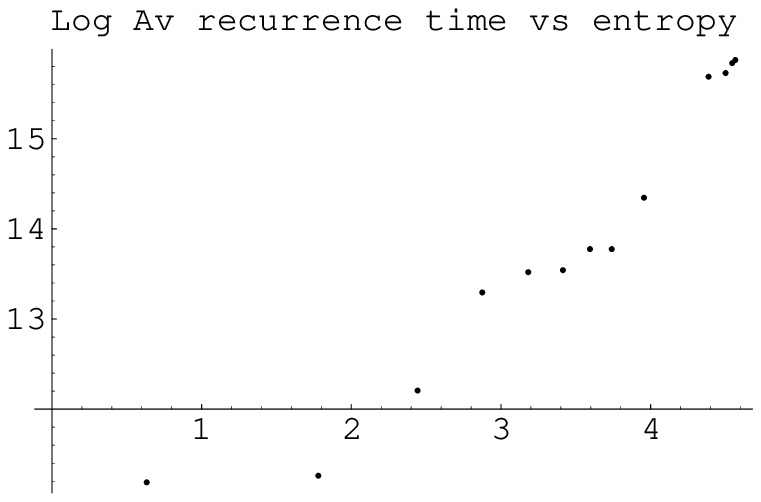}{Recurrence time as a power of $e^S$}

\bigbreak
As a final consideration, the long time average L from Equation \ref{lta}.  The functional
dependency of L is displayed in the final diagram (Figure \ref{gr12f}), 
\onefigure{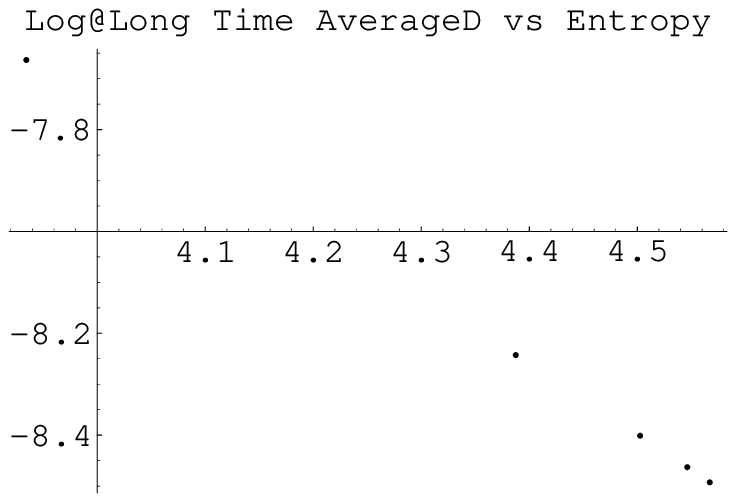}{Behavior of the log of the long time average as a function of the entropy} 
which shows the logarithm of the long time average as a function of the entropy
for a particular set of random matrices.  
This numerical result is consistent with the form $L \sim e^{-S}$, as previously stated.
\pagebreak
 
\end{document}